\documentclass[aps,prd,floatfix,groupedaddress,nofootinbib,superscriptaddress,twocolumn]{revtex4-2}

\usepackage{amsmath,amssymb,bm,color,float,graphicx,mathrsfs}
\usepackage[hidelinks]{hyperref}
\hypersetup{colorlinks=true,linkcolor=blue,citecolor=blue,urlcolor=blue}
\usepackage{Macro}

\begin{document}

\title{Exact CMB B-mode power spectrum from anisotropic cosmic birefringence}

\author{Toshiya Namikawa}
\affiliation{Center for Data-Driven Discovery, Kavli IPMU (WPI), UTIAS, The University of Tokyo, Kashiwa, 277-8583, Japan}

\date{\today}

\begin{abstract}
We calculate the cosmic microwave background (CMB) $B$-mode power spectrum resulting from anisotropic cosmic birefringence, without relying on the thin approximation of the last scattering surface. Specifically, we consider the influence of anisotropic cosmic birefringence arising from massless axion-like particles. Comparing our results to those obtained using the thin approximation, we observe a suppression in the amplitude of the $B$-mode power spectrum by approximately an order of magnitude at large angular scales ($\ell \lesssim 10$) and by a factor of two at small angular scales ($\ell \gtrsim 100$) when not employing the thin approximation. We also constrain the amplitude of the angular power spectrum of the scale-invariant anisotropic cosmic birefringence using the SPTpol $B$-mode power spectrum. We find that the amplitude is constrained as $A_{\rm CB}\times10^4=1.03^{+0.91}_{-0.97}\,(2\,\sigma)$. 
The numerical code is publicly available \href{https://github.com/toshiyan/biref-aniso-bb/tree/main}{here}.
\end{abstract} 



\maketitle

\section{Introduction} 

Cosmic birefringence --- a rotation of the linear polarization plane of photons
as they travel through space \cite{Ni:2007:cpr} --- is a key observational effect on the cosmic microwave background (CMB) as it provides a way to search for parity-violating physics in cosmology \cite{Komatsu:2022:review}. Recent analyses of the cross-correlation between the even-parity CMB $E$-modes and odd-parity CMB $B$-modes in the Planck polarization map suggest a tantalizing hint of cosmic birefringence \cite{Minami:2020:biref,Diego-Palazuelos:2022,Eskilt:2022:biref-freq,Eskilt:2022:biref-const, Eskilt:2023:cosmoglobe}. 
Cosmic birefringence has also been explored with non-CMB sources, including the Crab Nebula by \textsc{POLARBEAR} which recently placed a median $2\sigma$ upper bound of polarization oscillation amplitude $A<0.065$\,deg over the oscillation frequencies from $0.75$\,year$^{-1}$ to $0.66$\,hour$^{-1}$ \cite{POLARBEAR:2024vel}. 

Cosmic birefringence can be induced by a pseudoscalar field, such as axionlike particles (ALPs), coupled with electromagnetic fields through the Chern-Simons term, $\mathcal{L}\supset -g_{\phi\gamma}\phi F^{\mu\nu}\tilde{F}{\mu\nu}/4$, where $g{\phi\gamma}$ represents the coupling constant, $\phi$ denotes an ALP field, $F^{\mu\nu}$ stands for the electromagnetic field tensor, and $\tilde{F}_{\mu\nu}$ is its dual.
Multiple works have explored cosmic birefringence induced by the ALP field of dark energy \cite{Carroll:1998:DE,Panda:2010,Fujita:2020aqt,Fujita:2020ecn,Choi:2021aze,Obata:2021,Gasparotto:2022uqo,Galaverni:2023}, early dark energy \cite{Fujita:2020ecn,Murai:2022:EDE,Eskilt:2023:EDE}, dark matter \cite{Finelli:2009,Liu:2016dcg,Fedderke:2019:biref}, and by topological defects \cite{Takahashi:2020tqv,Kitajima:2022jzz,Jain:2022jrp,Gonzalez:2022mcx}, as well as by potential signatures of quantum gravity \cite{Myers:2003fd,Balaji:2003sw,Arvanitaki:2009fg}. Anticipated advancements in CMB experiments, including BICEP \cite{Cornelison:2022:BICEP3,BICEPArray}, Simons Array \cite{SimonsArray}, Simons Observatory \cite{SimonsObservatory}, CMB-S4 \cite{CMBS4}, and LiteBIRD \cite{LiteBIRD}, are projected to substantially reduce polarization noise, thus enhancing sensitivity to cosmic birefringence signals.

Several recent studies have shown that the time evolution of the pseudoscalar fields during recombination and reionization significantly changes the $EB$ power spectrum \cite{Finelli:2009,Sigl:2018:biref-sup,Sherwin:2021:biref,Nakatsuka:2022,Murai:2022:EDE,Eskilt:2023:EDE,Liu:2006,Lee:2013:const,Gubitosi:2014:biref-time}. Measuring the spectral shape of the $EB$ power spectrum will provide tomographic information on the pseudoscalar fields. We can also constrain the late-time evolution by observing the polarized Sunyaev-Zel'dovich effect \cite{Lee:2022:pSZ-biref,Namikawa:2023:pSZ} and galaxies \cite{Carroll:1997:radio,Yin:2024:galaxy}. This tomographic method can avoid the degeneracies with the instrumental miscalibration angle \cite{QUaD:2008ado,Komatsu:2010:WMAP7,B1rot,P16:rot} and half-wave plate nonidealities \cite{Monelli:2022:HWP}. 

If the ALP fields fluctuate, $\delta\phi$, which has a spatial variation, the polarization rotation will be anisotropic (i.e., have different values in different directions in the sky); indeed, anisotropies in the cosmic birefringence are produced by many of the types of beyond-the-Standard-Model physics (see, e.g., \cite{Carroll:1998:DE,Lue:1999:biref-EB,Caldwell:2011,Lee:2015,Leon:2017,Yin:2023:biref,Ferreira:2023}).
In this paper, we investigate the impact of the time evolution of $\delta\phi$ on the calculation of anisotropic cosmic birefringence. Specifically, we focus on the CMB $B$-mode power spectrum which is most sensitive to anisotropic cosmic birefringence among the CMB two-point correlations.

The $B$-mode power spectrum arising from anisotropic cosmic birefringence has often been computed using the approximation of the thin last-scattering surface (LSS). This approximation ignores the time evolution of $\delta\phi$ during the recombination (and reionization) epoch to simplify the $B$-mode power spectrum \cite{Li:2008tma,Li:2013,Zhao:2014,Lee:2015,Lee:2016,Liu:2016dcg,Pogosian:2019:forecast,Cai:2021:classrot,CoS:2021}. Our main objective is to demonstrate the significance of considering the thickness of the LSS when computing the $B$-mode power spectrum. It's worth noting that only a few studies have incorporated the time evolution of $\delta\phi$ to derive the $B$-mode power spectrum without the thin LSS approximation. Pospelov et al. (2009) \cite{Pospelov:2009} (hereafter, P09) derived the expression for the $B$-mode power spectrum from time-varying anisotropic cosmic birefringence and numerically computed the power spectrum. Greco et al. (2024) \cite{Greco:2024} also derived the expression for the $B$-mode power spectrum but did not numerically compute it due to numerical challenges. We further compare our results with the previous numerical computations by P09.

This paper is organized as follows. In section \ref{sec:biref}, we review cosmic birefringence and CMB $B$-mode power spectrum. In section \ref{sec:exact}, we derive the $B$-mode power spectrum induced by anisotropic cosmic birefringence without the thin LSS approximation and evaluate the $B$-mode power spectrum numerically. Section \ref{sec:summary} is devoted to summary and conclusion.

\section{Cosmic birefringence} \label{sec:biref}

\subsection{Cosmic birefringence from massless scalar fields}

We first briefly review cosmic birefringence. The presence of the pseudoscalar fields rotates the polarization plane of CMB photons. This rotation angle in general depends on the line-of-sight direction, $\hatn$. The rotation angle for photons emitted at conformal time $\eta$ in a line-of-sight direction $\hatn$ is given by \cite{Harari:1992,Carroll:1989:rot,Carroll&Field:1991} 
\footnote{We here ignore the gravitational lensing effect which changes the trajectory of the CMB photons. Even if we include the lensing effect, the isotropic cosmic birefringence remains unchanged \cite{Namikawa:2021:mode,Naokawa:2023}. On the other hand, the anisotropic cosmic birefringence is modified since the rotation angle depends on $\delta\phi$ at a slightly different position from $\chi\hatn$. This remapping effect would be, however, negligible in the anisotropic cosmic birefringence if its angular power spectrum is red; the lensing remapping effect changes any red spectrum only slightly \cite{Lewis:2006:review}. Since the model for the anisotropic cosmic birefringence in this paper has a steep red spectrum, we ignore the lensing effect on the birefringence angle throughout the paper.}
\al{
    \beta(\eta,\hatn) = \frac{g}{2}[\bar{\phi}(\eta_0)-\bar{\phi}(\eta)-\delta\phi(\eta,\chi\hatn)]
    \,. 
}
Here, $\chi=\eta_0-\eta$ is the comoving distance of the source. 
The rotation angle is decomposed into the isotropic and anisotropic components. Anisotropies of the rotation angle are given by
\al{
    \alpha(\eta,\hatn) \equiv -\frac{g_\phi}{2}\delta\phi(\eta,\chi\hatn)
    \,. 
}

Throughout this paper, we consider the massless pseudoscalar field as the source of the anisotropic cosmic birefringence since the time evolution of the pseudoscalar field is well known and the model can produce a large signal for the anisotropic cosmic birefringence.
In this case, the isotropic components become zero and the rotation angle is given by the anisotropic component, $\beta(\eta,\hatn)=\alpha(\eta,\hatn)$ \cite{Caldwell:2011}. 
The evolution equation for the massless pseudoscalar field fluctuations is given by \cite{Caldwell:2011}: 
\al{
    \D{^2\delta\phi}{\eta^2} + 2\mC{H}\D{\delta\phi}{\eta} + k^2\delta\phi = 0
    \,, \label{Eq:dphi:EoM}
}
where $\mC{H}=({\rm d}a/{\rm d}\eta)/a$ with $a$ being the scale factor. We introduce the transfer function to characterize the evolution of the fluctuations: 
\al{
    \delta\phi(\eta,\bm{k}) = T(k,\eta)\delta\phi_{\rm ini}(\bm{k})
    \,. 
}
For the massless pseudoscalar case, during the matter domination epoch, we have
\al{
    T(k,\eta) &= 3\frac{j_1(k\eta)}{k\eta}
    \,. 
}
The transfer function satisfies $T(k,\eta)\to 1$ at the super-horizon scale. We define the correlation of the pseudoscalar fluctuations as \cite{Greco:2022:aniso-biref-tomography}
\al{
    \ave{\delta\phi^*_{\rm ini}(\bm{k})\delta\phi_{\rm ini}(\bm{k}')} = \frac{2\pi^2}{k^3}{\cal P}_{\phi}(k) (2\pi)^3\delta^{(3)}(\bm{k}-\bm{k}')
    \,. 
}
The dimensionless primordial power spectrum for the pseudoscalar fields originated from the initial vacuum fluctuations is given by \cite{Caldwell:2011}:
\al{
    {\cal P}_\phi(k) &= \left(\frac{H_I}{2\pi}\right)^2 
    \,, 
}
where $H_I$ is the expansion rate during inflation.

\subsection{B-mode power spectrum with the thin LSS approximation}

Here, we introduce the $B$-mode power spectrum with the thin LSS approximation. The polarization plane of the CMB photon from the LSS is rotated during propagation, and the Stokes $Q$ and $U$ polarization map is distorted as
\al{
    [Q\pm \iu U](\hatn) = [\bar{Q}\pm \iu \bar{U}](\hatn) \E^{\pm 2\iu\alpha(\eta_*,\hatn)}
    \,, \label{Eq:QU:rot}
}
where the bar denotes the CMB anisotropies without the rotation. 
The $E$- and $B$-mode polarization are then obtained by \cite{Zaldarriaga:1996xe}
\al{
	E_{\l m} \pm \iu B_{\l m} = -\Int{2}{\hatn}{} {}_{\pm 2}Y_{\l m}^*(\hatn) [Q\pm\iu U](\hatn) 
	\,, \label{Eq:EB-def}
}
where ${}_{\pm 2}Y_{\l m}(\hatn)$ is the spin-$2$ spherical harmonics. Similarly with the spin-$0$ (scalar) spherical harmonics, $Y_{\l m}(\hatn)$, the harmonic coefficients of the rotation angle is defined as
\al{
	\alpha_{LM}(\eta_*) &= \Int{2}{\hatn}{} Y_{LM}^*(\hatn) \alpha (\eta_*,\hatn) 
	\,.
}
Expanding the right-hand side of \eq{Eq:QU:rot} to first order in
the rotation angle and transforming the Stokes $Q/U$ parameters
to $E/B$ modes using \eq{Eq:EB-def}, we obtain the distorted $B$ modes by the anisotropic cosmic birefringence as
\al{
    B_{\l m} &= \bar{B}_{\l m} + \sum_{LM\l'm'}(-1)^m\Wjm{\l}{L}{\l'}{-m}{M}{m'}
    \notag \\
    &\times \alpha_{LM}(\eta_*)\left(-W^-_{\l L\l'}\bar{E}_{\l'm'}+W^+_{\l L\l'}\bar{B}_{\l'm'}\right)
    \,. 
}
Here, we define 
\al{
    W^{\pm}_{\l L\l'} &= \pm 2\iu \zeta^\pm p^\mp_{\l L\l'} \sqrt{\frac{(2\l+1)(2L+1)(2\l'+1)}{4\pi}}\Wjm{\l}{L}{\l'}{2}{0}{-2}
    \,, 
}
with $\zeta^+=1$, $\zeta^-=\iu$, and $p^{\mp}_{\l L\l'}=[1\mp(-1)^{\l+L+\l'}]/2$ is the result of the parity symmetry. The $B$-mode power spectrum with the thin LSS approximation is given by (e.g. \cite{Li:2008tma})
\al{
    C_\l^{BB} 
    &= \ave{|B_{\l m}|^2} 
    = \sum_{\l'L}\frac{1}{2\l+1}C_L^{\alpha\alpha}(\eta_*,\eta_*)
    \notag \\
    &\times [|W_{\l L\l'}^-|^2 C_{\l'}^{\bar{E}\bar{E}} + |W_{\l L\l'}^+|^2 C_{\l'}^{\bar{B}\bar{B}}] 
    \notag \\
    &= 4\sum_{\l'L}\frac{(2\l'+1)(2L+1)}{4\pi} \Wjm{\l}{L}{\l'}{2}{0}{-2}^2
    \notag \\
    &\times C_L^{\alpha\alpha}(\eta_*,\eta_*)[p^+_{\l L\l'}C_{\l'}^{\bar{E}\bar{E}}+p^-_{\l L\l'}C_{\l'}^{\bar{B}\bar{B}}] 
    \,. \label{Eq:BB:thin}
}
Here, we define the angular power spectrum of the anisotropic rotation fields as \cite{Greco:2022:aniso-biref-tomography} 
\al{
    C_L^{\alpha\alpha}(\eta,\eta') 
    &= \ave{\alpha_{LM}(\eta)\alpha^*_{LM}(\eta')}
    \notag \\
    &= 4\pi \Int{}{\ln k}{} {\cal P}_\phi(k) u_L(k,\eta)u_L(k,\eta')
    \,, \label{Eq:Claa}
}
where we define 
\al{
    u_L(k,\eta) = \frac{g_\phi}{2} j_L(k(\eta_0-\eta))T(k,\eta)
    \,. 
}
At large-angular scales, $L\alt 100$, the transfer function is approximately unity, and the angular power spectrum of the anisotropic birefringence angle is given by \cite{Caldwell:2011}: 
\al{
    C_L^{\alpha\alpha} \simeq \frac{2\pi}{L(L+1)}\left(\frac{g_{\phi}}{2}\right)^2\left(\frac{H_I}{2\pi}\right)^2 \equiv \frac{2\pi}{L(L+1)} A_{\rm CB} 
    \,, 
}
where we introduce a constant, $A_{\rm CB}$, and use
\al{
    \INT{}{\ln x}{}{0}{\infty} j^2_L(x) = \frac{1}{2L(L+1)} 
    \,. 
}

\subsection{Other contributions}

\begin{figure}
    \centering
    \includegraphics[width=80mm]{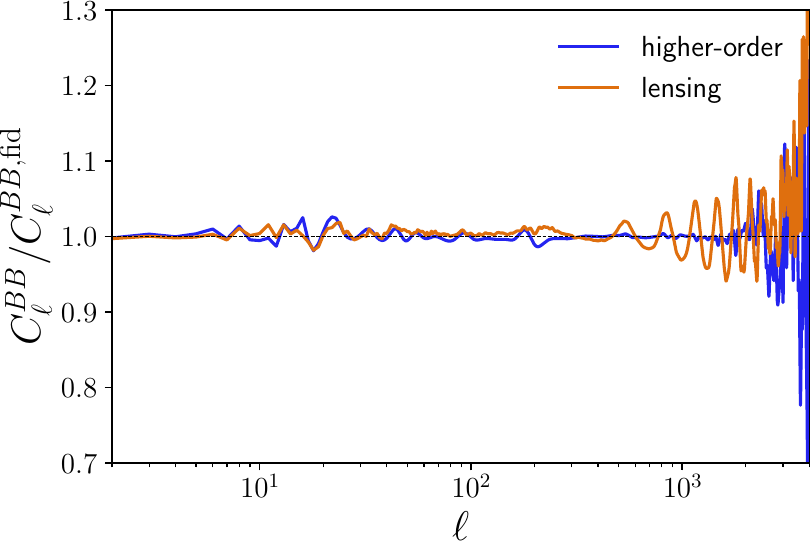}
    \caption{Ratio of the $B$-mode power spectra with and without the higher-order terms, $\mC{O}[(C^{\alpha\alpha}_L)^2]$ (``higher-order'') and with and without the lensing effect (``lensing'').}
    \label{fig:higher_lensing}
\end{figure}

In the above calculation, we made some simplifications. For example, we ignored the contributions from the higher-order terms of $\mC{O}[(C_L^{\alpha\alpha})^2]$ to the $B$-mode power spectrum. We also ignore the lensing contributions. 

To see the impact of the higher-order terms and lensing, we compare the cases with and without the higher-order terms or the lensing effect using the public software package \href{https://github.com/catketchup/class_rot}{{\tt class\_rot}} developed by Cai \& Guan (2021) \cite{Cai:2021:classrot} with the thin LSS approximation. We find that these contributions are less than a few percent at $\l\alt 1000$ and we can approximate $B$-mode power spectrum with the perturbed spectrum without lensing. On the other hand, at very small scales, $\l\agt 3000$, both contributions become important.

\section{B-mode power spectrum from anisotropic cosmic birefringence} \label{sec:exact}

Here, we derive the expression for the $B$-mode power spectra from time-varying anisotropic cosmic birefringence without the thin LSS approximation.

\subsection{B-mode power spectrum without the thin LSS approximation}

We follow Ref.~\cite{Pogosian:2011:thick} where they derive the $B$-mode power spectrum from the Faraday rotation due to the primordial magnetic fields without the thin LSS approximation. The derivation is based on the total angular momentum method of Ref.~\cite{Hu:1997:TAM}. 

The evolution equation for the $\bm{q}$ Fourier mode of $P_\pm=Q\pm \iu U$ is given by \cite{Pogosian:2011:thick,Nakatsuka:2022}
\al{
    &\D{P_\pm}{\eta}(\eta,\bm{q},\hatn) + \iu q\mu P_\pm(\eta,\bm{q},\hatn) 
    \notag \\
    &= \dot{\tau}\left[-P_\pm(\eta,\bm{q},\hatn)+\sqrt{6}P^{(0)}(\eta,\bm{q})\sqrt{\frac{4\pi}{5}}{}_{\pm2}Y_{20}(\hatn)\right] 
    \notag \\
    &\mp\iu \D{\alpha}{\eta}(\eta,\hatn) P_\pm(\eta,\bm{q},\hatn)
    \,. 
}
Here, $\mu=\hatn\cdot\hat{\bm{q}}$, and we define the differential CMB optical depth, $\dot{\tau}\equiv an_{\rm e}\sigma_{\rm T}$, with $n_{\rm e}$ and $\sigma_{\rm T}$ being the electron number density and the Thomson-scattering cross-section, respectively. The quantity, $P^{(0)}=(\Theta_2-\sqrt{6}E_2)/10$, is the polarization source with $\Theta_2$ and $E_2$ being the quadrupole of temperature and $E$-mode polarization from the scalar perturbations \cite{Hu:1997:TAM}. The above equation has the following solution: 
\al{
    P_\pm(\bm{q},\hatn) &= - \INT{}{\eta}{}{0}{\eta_0} g_{\rm v}(\eta) \sqrt{6}P^{(0)}(\eta,\bm{q})\left[1\pm2\iu\alpha(\eta,\hatn)\right]
    \notag \\
    &\times \sum_\l(-\iu)^\l\sqrt{4\pi(2\l+1)}\epsilon^{(0)}_\l(q(\eta_0-\eta)){}_{\pm2}Y_{\l0}(\hatn)
    \notag \\
    &+ \mC{O}(\alpha^2)
    \,, \label{Eq:P:solution}
}
where $g_{\rm v}(\eta)$ is the visibility function and
\al{
    \epsilon_\l^{(0)}(x) = \sqrt{\frac{3}{8}\frac{(\l+2)!}{(\l-2)!}}\frac{j_\l(x)}{x^2}
    \,. 
}

Next, we use the total angular momentum method to derive the expression for the $B$-mode power spectrum. We decompose the polarization fields as 
\al{
    P_{\pm}(\bm{q},\hatn) &= \sum_\l(-\iu)^\l\sqrt{\frac{4\pi}{2\l+1}}
    \notag \\ 
    &\times \sum_{m=-\l}^\l(E_\l^{(m)}(\bm{q})\pm\iu B_\l^{(m)}(\bm{q})){}_{\pm 2}Y_{\l m}(\hatn)
    \,, 
}
with $\hat{\bm{q}}=\hat{\bm{z}}$. Note that $B_\l^{(m)}(\bm{q})$ vanishes in the absence of cosmic birefringence, and the scalar perturbations alone lead to $m=0$. We then invert the above equation with the spin-2 spherical harmonics and obtain: 
\al{
    B_\l^{(m)}(\bm{q}) &= \iu^\l\sqrt{\frac{2\l+1}{4\pi}}
    \sum_{\l_1}(-\iu)^{\l_1}\sqrt{4\pi(2\l_1+1)}\epsilon^{(0)}_{\l_1}
    \notag \\
    &\times \Int{2}{\hatn}{}[{}_2Y_{\l_10}(\hatn){}_2Y_{\l m}^*(\hatn)+{}_{-2}Y_{\l_10}(\hatn){}_{-2}Y_{\l m}^*(\hatn)]
    \notag \\
    &\times \INT{}{\eta}{}{0}{\eta_0} g_{\rm v}(\eta)\sqrt{6}P^{(0)}(\bm{q},\eta)\alpha(\eta,\hatn)
    \,, \label{Eq:B-mode}
}
where we use \eq{Eq:P:solution}. The $B$-mode power spectrum is defined as \cite{Hu:1997:TAM}
\al{
    C_\l^{BB} = \frac{4\pi}{(2\l+1)^2}\Int{3}{\bm{q}}{(2\pi)^3}\sum_{m=-\l}^\l\ave{B_\l^{(m)*}(\bm{q})B_\l^{(m)}(\bm{q})}
    \,. \label{Eq:ClBB:def}
}
Substituting \eq{Eq:B-mode} into \eq{Eq:ClBB:def}, the $B$-mode power spectrum in the presence of the time-varying anisotropic cosmic birefringence is given by 
\begin{widetext}
\al{
    C^{BB}_\l &= \frac{4\pi}{2\l+1}\sum_{m=-\l}^\l\Int{}{\ln q}{}{\cal P}_{\cal R}(q)
    \sum_{\l_1\l_2}\iu^{-\l_1+\l_2}\sqrt{(2\l_1+1)(2\l_2+1)}
    \notag \\
    &\times 
    \Int{2}{\hatn}{}[{}_2Y_{\l_10}^*(\hatn){}_2Y_{\l m}(\hatn)+{}_{-2}Y_{\l_10}^*(\hatn){}_{-2}Y_{\l m}(\hatn)]
    \Int{2}{\hatn'}{}[{}_2Y_{\l_20}(\hatn'){}_2Y_{\l m}^*(\hatn')+{}_{-2}Y_{\l_20}(\hatn'){}_{-2}Y_{\l m}^*(\hatn')]
    \notag \\
    &\times \INT{}{\eta}{}{0}{\eta_0}\INT{}{\eta'}{}{0}{\eta_0} s_{\l_1}(q,\eta) s_{\l_2}(q,\eta')
    \ave{\alpha(\eta,\hatn)\alpha(\eta',\hatn')}
    \,, \label{Eq:ClBB}
}
\end{widetext}
where we relate the correlation of the polarization source to the primordial curvature power spectrum as
\al{
    \ave{[P^{(0)}(\bm{q},\eta)]^*P^{(0)}(\bm{q},\eta')} = \frac{2\pi^2}{q^3}{\cal P}_{\cal R}(q)S(q,\eta)S(q,\eta')
    \,, 
}
and define a projected polarization source function
\al{
    s_\l(q,\eta) \equiv g_{\rm v}(\eta)\sqrt{6}S(\eta,q)\epsilon^{(0)}_\l(q(\eta_0-\eta))
    \,. 
}
Note that in deriving Eq.~\eqref{Eq:ClBB} we assume that the source of the anisotropic cosmic birefringence does not correlate with the curvature perturbations. This assumption is valid for the massless pseudoscalar field model considered in this paper but does not hold in general. 

We use the following expression of the two-point correlation function of the anisotropic cosmic birefringence: 
\al{
    &\ave{\alpha(\eta,\hatn)\alpha(\eta',\hatn')} 
    = \sum_{LM} Y^*_{LM}(\hatn)Y_{LM}(\hatn') C_L^{\alpha\alpha}(\eta,\eta')
    \,. \label{Eq:alpha-correlation}
}
Substituting \eq{Eq:alpha-correlation} into \eq{Eq:ClBB}, and employing the formulas for the Wigner $3j$ symbols in Appendix \ref{sec:wigner3j}, we finally obtain the $B$-mode power spectrum as
\al{
    C_\l^{BB} = 4\sum_{\l'L}p^+_{\l L\l'}\frac{(2\l'+1)(2L+1)}{4\pi}\Wjm{\l}{L}{\l'}{2}{0}{-2}^2{\cal C}^{EE}_{\l',L}
    \,, \label{Eq:exactBB}
}
where the distorted $E$-mode power spectrum is defined as
\al{
    {\cal C}^{EE}_{\l',L} \equiv 4\pi \Int{}{\ln k}{} {\cal P}_{\phi}(k) {\cal C}^{EE}_{\l',L}(k)
    \,. \label{Eq:total-dist-E}
}
Here, we define the $E$-mode power spectrum for each $k$ as
\al{
    {\cal C}^{EE}_{\l',L}(k) \equiv 4\pi \Int{}{\ln q}{} {\cal P}_{\cal R}(q) \left[\Delta_{\l',L}(q,k)\right]^2 
    \,, \label{Eq:total-dist-E-k}
}
with
\al{
    \Delta_{\l',L}(q,k) = \INT{}{\eta}{}{0}{\eta_0} s_{\l'}(q,\eta) u_L(k,\eta)
    \,. 
}

Note that the thin LSS approximation leads to
\al{
    \Delta_{\l',L}(q,k) \simeq u_L(k,\eta_*)\INT{}{\eta}{}{0}{\eta_0} s_{\l'}(q,\eta) 
    \,, 
}
and we obtain
\al{
    {\cal C}^{EE}_{\l',L} &\simeq 4\pi \Int{}{\ln k}{} {\cal P}_{\phi}(k) [u_L(k,\eta_*)]^2 
    \notag \\
    &\times 4\pi \Int{}{\ln q}{} {\cal P}_{\cal R}(q)\left[\INT{}{\eta}{}{0}{\eta_0} s_{\l'}(q,\eta) \right]^2 
    \notag \\
    &= C_L^{\alpha\alpha}(\eta_*,\eta_*)C_{\l'}^{\bar{E}\bar{E}}
    \,, \label{Eq:total-dist-E-k:thin}
}
where we use \eq{Eq:Claa} and the expression for the primordial $E$-mode power spectrum:
\al{
    C_{\l'}^{\bar{E}\bar{E}} = 4\pi \Int{}{\ln q}{} {\cal P}_{\cal R}(q)\left[\INT{}{\eta}{}{0}{\eta_0} s_{\l'}(q,\eta)\right]^2
    \,. 
}
Substituting \eq{Eq:total-dist-E-k:thin} into \eq{Eq:exactBB} coincides with the expression for the birefringence-induced $B$-mode power spectrum with the thin LSS approximation of \eq{Eq:BB:thin} but without the intrinsic $B$-mode power spectrum.

\subsection{Impact of the thin LSS approximation}

\begin{figure*}
    \centering
    \includegraphics[width=80mm]{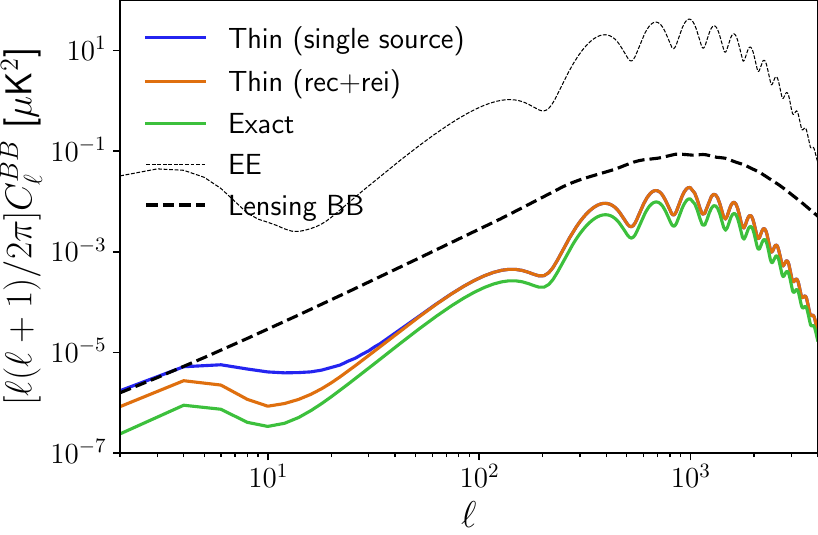}
    \includegraphics[width=80mm]{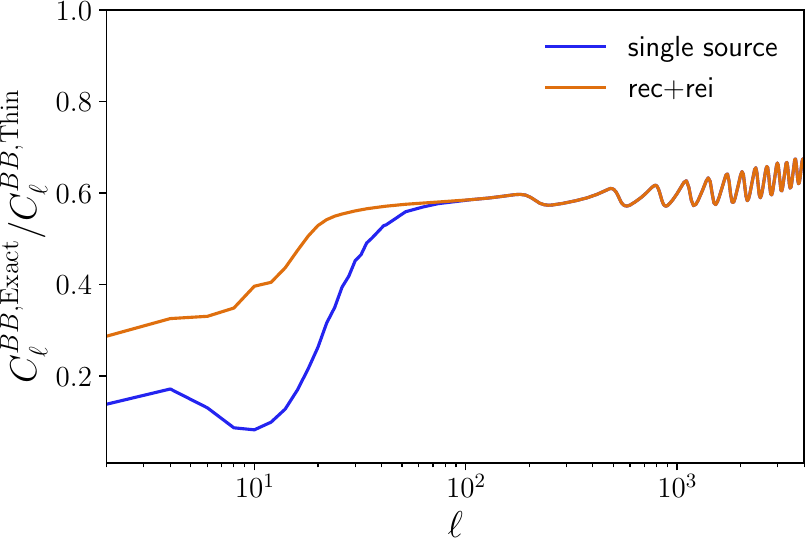}
    \caption{Comparison of the $B$-mode power spectra with and without the thin approximation denoted as ``exact'' and ``thin'', respectively ({\it Left}), and their ratio ({\it Right}). For the thin case, we also show the two-source approximation where the $B$-mode power spectrum is the sum of the contributions from the recombination and reionization epochs. In {\it Left}, as a reference, we plot the $E$-mode power spectrum in a black solid line and the lensing-induced $B$-mode power spectrum in a black dashed line.}
    \label{fig:BB}
\end{figure*}

\begin{figure*}
    \centering
    \includegraphics[width=80mm]{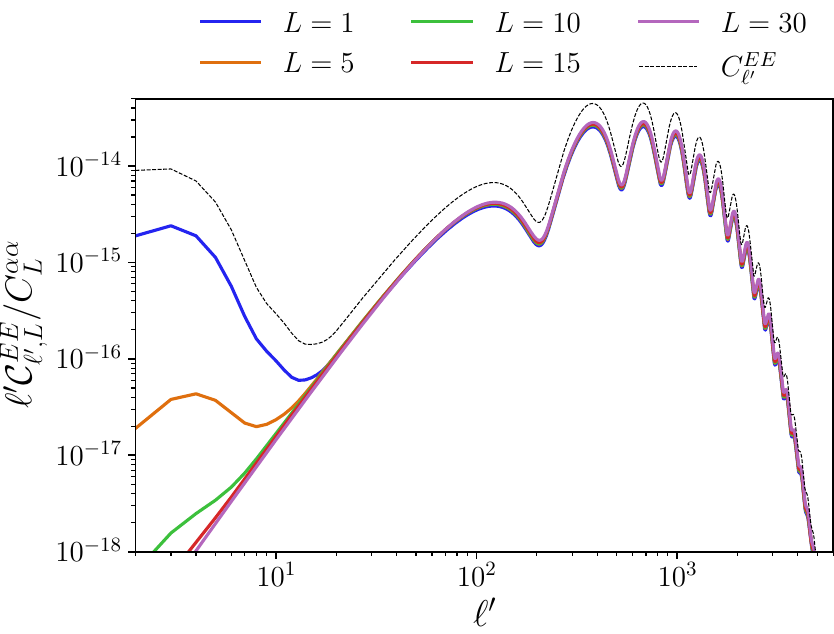}
    \includegraphics[width=80mm]{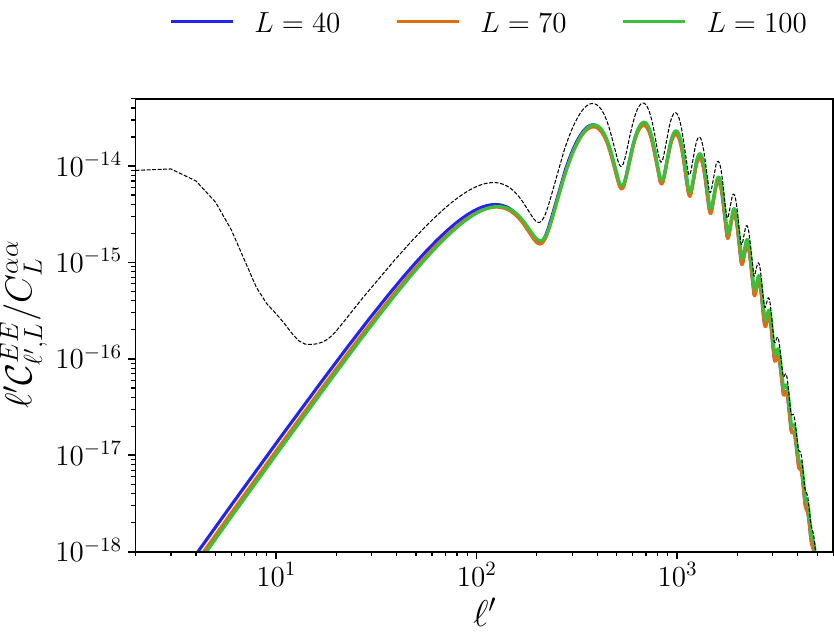}
    \caption{The distorted $E$-mode power spectrum defined in \eq{Eq:total-dist-E} divided by the anisotropic birefringence power spectrum. Since ${\cal C}^{EE}_{\l',L}/C_L^{\alpha\alpha}\to C_{\l'}^{EE}$ in the thin approximation, we also plot the $E$-mode power spectrum for comparison.}
    \label{fig:calC}
\end{figure*}

Here, we compare the $B$-mode power spectra with and without the thin LSS approximation. 
To compute the exact $B$-mode power spectrum, we first compute the distorted $E$-mode power spectrum per wavenumber given in Eq.~\eqref{Eq:total-dist-E-k} by modifying the Boltzmann code \href{https://lesgourg.github.io/class_public/class.html}{{\tt CLASS}} \cite{CLASS}. We then numerically integrate ${\cal C}^{EE}_{\l',L}(k)$ to obtain ${\cal C}^{EE}_{\l',L}$ from Eq.~\eqref{Eq:total-dist-E}. We finally obtain the $B$-mode power spectrum from Eq.~\eqref{Eq:exactBB} where we employ \href{https://github.com/xzackli/WignerFamilies.jl}{{\tt WignerFamilies}} to compute the Wigner 3j symbols. We choose the amplitude of ${\cal P}_\phi(k)$ so that $A_{\rm CB}=3\times 10^{-5}$ which corresponds to the current bound from observations \cite{Namikawa:2020:biref,SPT:2020:brief,Zagatti:2024}. 

Figure~\ref{fig:BB} shows the $B$-mode power spectra with and without the thin LSS approximation. We ignore the primordial $B$-mode power spectrum in \eq{Eq:BB:thin}. The amplitude of the $B$-mode power spectrum is suppressed by an order of magnitude at $\l<10$, and by approximately a factor of two at $\l>100$. We also show the $B$-mode power spectrum with the thin LSS approximation, \eqref{Eq:BB:thin}, but separating the contribution from the recombination and reionization, i.e., 
\al{
    &C_\l^{BB,\rm rec} + C_\l^{BB,\rm rei} = 4\sum_{\l'L}p^+_{\l L\l'}\frac{(2\l'+1)(2L+1)}{4\pi} 
    \notag \\
    &\qquad \times \Wjm{\l}{L}{\l'}{2}{0}{-2}^2
    \sum_{x={\rm rei},{\rm rec}}C_L^{\alpha\alpha}(\eta_x,\eta_x)C_{\l'}^{\bar{E}\bar{E},x} 
    \,, \label{Eq:BB:rei+rec} 
}
where we approximate that $C_{\l'}^{\bar{E}\bar{E},{\rm rei}}$ and $C_{\l'}^{\bar{E}\bar{E},{\rm rec}}$ are obtained from the $E$-mode power spectrum at $\l'\leq 10$ and $10<\l'$, respectively. The exact $B$-mode power spectrum is still different from this two-source approximation. 

Figure~\ref{fig:calC} shows the ratio, ${\cal C}^{EE}_{\l',L}/C_L^{\alpha\alpha}$ for different $L$. In the thin LSS limit, this ratio becomes equivalent to the $E$-mode power spectrum. The ratio is substantially suppressed at low $\l'$. If the $B$-mode multipole is $\l\alt 10$ in \eq{Eq:exactBB}, the contribution to the $B$-mode power spectrum comes from $\l'\simeq L$ due to the property of the Wigner $3j$ symbol. Since the rotation field is significant at low-$L$, the significant reduction of the amplitude in the low-$\l$ $B$-mode power spectrum in Fig.~\ref{fig:BB} would be the result of the substantial suppression appeared in ${\cal C}^{EE}_{\l',L}/C_L^{\alpha\alpha}$ at low-$\l'$. On the other hand, the $B$-modes at high-$\l$ are also produced by high-$\l'$ $E$-modes, and the suppression in the $B$-mode power spectrum becomes mild.

\subsection{Comparison with previous works}

Our results show that the $B$-mode power spectrum is significantly suppressed if we appropriately include the time evolution of the pseudoscalar fields. On the other hand, P09 \cite{Pospelov:2009} shows that the $B$-mode power spectrum at low-$\l$ is rather enhanced significantly. We here discuss possible sources of the difference between our results and the previous attempt to compute the $B$-mode power spectrum with time-varying anisotropic cosmic birefringence by P09. 

P09 derives the $B$-mode power spectrum by extending the formula of Ref.~\cite{Zaldarriaga:1996xe}. Choosing a Fourier wavenumber of a single scalar perturbation mode being parallel to the $z$ axis, cosmic birefringence induces the Stokes $U$ polarization as $U(\hatn,\eta)=2\alpha(\hatn,\eta)\bar{Q}(\hatn,\eta)$ at conformal time $\eta$ at the lowest order of $\alpha$. We then obtain the $B$-mode as
\al{
    B_{\l m} &= -\sqrt{\frac{(\l-2)!}{(\l+2)!}}\Int{2}{\hatn}{}Y_{\l m}^*(\hatn) \frac{\eth^2+\bar{\eth}^2}{2}U(\hatn) 
    \notag \\
    &= \frac{3}{2}\sqrt{\frac{(\l-2)!}{(\l+2)!}}\Int{2}{\hatn}{}Y^*_{\l m}(\hatn) \Int{3}{\bm{q}}{} 
    \notag \\
    &\times \INT{}{\eta}{}{0}{\eta_0} \frac{\eth^2+\bar{\eth}^2}{2} \E^{\iu x\mu}\alpha(\hatn,\eta) g_{\rm v}(\eta)\Pi(q,\eta)\xi(\bm{q})
    \,, 
}
where $\eth$ and $\bar{\eth}$ are the spin rising and lowering operators, respectively, $x=q(\eta_0-\eta)$, $\Pi(q,\eta)$ is the polarization source \cite{Zaldarriaga:1996xe}, and $\xi(\bm{q})$ is the primordial fluctuations. 
P09 replaces the spin operators as $(\eth^2+\bar{\eth}^2)/2=m^2-(1+\pd_x^2)^2x^2$ and obtains the $B$-modes induced by cosmic birefringence as
\al{
    B_{\l m} &= \frac{3}{2}\sqrt{\frac{(\l-2)!}{(\l+2)!}}\Int{2}{\hatn}{}\Int{3}{\bm{q}}{}\Int{3}{\bm{k}}{} Y^*_{\l m}(\hatn) 
    \notag \\
    &\times \INT{}{\eta}{}{0}{\eta_0} [m^2-(1+\pd_x^2)^2x^2]\E^{\iu x\mu+\iu y\nu}
    \notag \\
    &\times g_{\rm v}(\eta)\Pi(q,\eta)T(k,\eta)\xi(\bm{q})\frac{g_{\phi}}{2}\delta\phi(\bm{k})
    \,, 
}
where $y=k(\eta_0-\eta)$ and $\nu=\hatn\cdot\hat{\bm{k}}$. 
The above expression, however, ignores the operation of the spin operator to $\alpha(\hatn,\eta)$. Thus, the expression for the $B$-mode power spectrum does not reduce to the $B$-mode power spectrum derived in the literature even in the thin LSS approximation. This discrepancy would be a source of the difference in the behavior of the $B$-mode power spectrum at large angular scales.

\subsection{Constraints on anisotropic cosmic birefringence with B-mode power spectrum}

\begin{figure*}
    \centering
    \includegraphics[width=85mm]{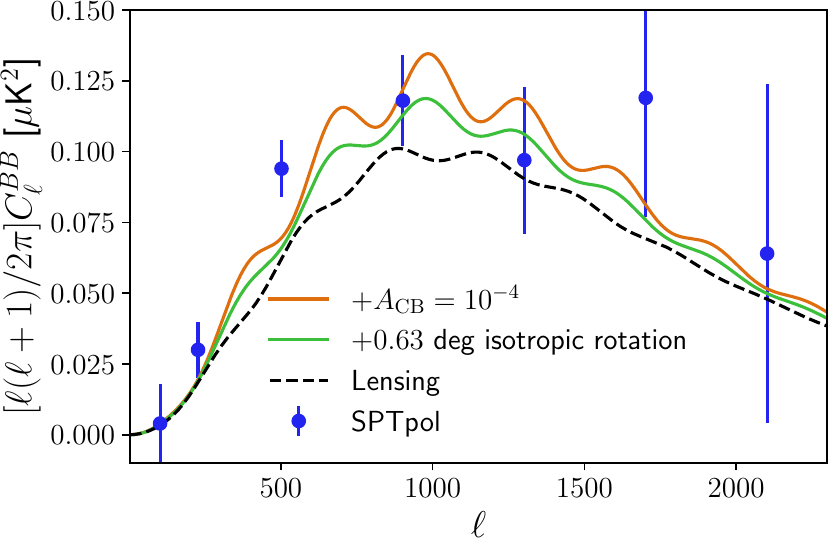}
    \includegraphics[width=85mm]{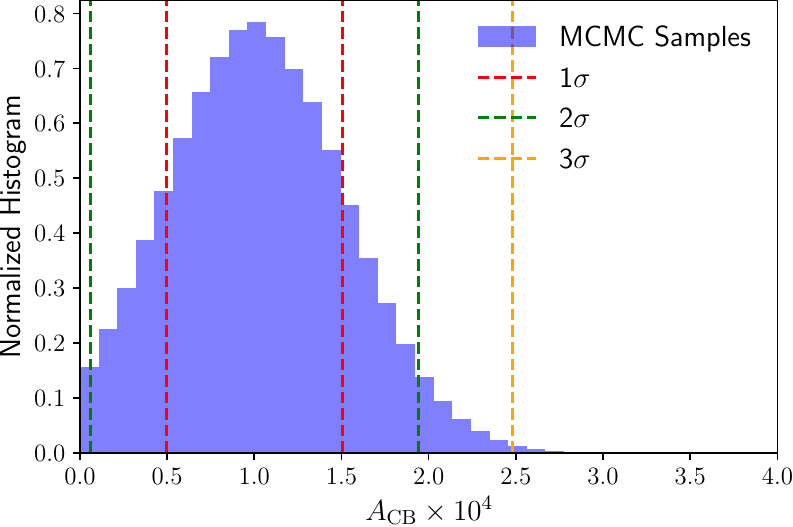}
    \caption{
    {\it Left}: The angular power spectrum from the anisotropic cosmic birefringence and lensing with parameters corresponding to the best-fit values (orange solid), the lensing-only power spectrum (black dashed), and the bandpower of the SPTpol $B$-mode power spectrum. For reference, we also show the $B$-mode power spectrum from the isotropic polarization rotation of $0.63\,$deg with lensing (green solid) where the angle is estimated from Ref.~\cite{SPT:2020:brief}.
    {\it Right}: The posterior distribution of $A_{\rm CB}$ with $1$, $2$, and $3$\,$\sigma$ regions. 
    }
    \label{fig:const}
\end{figure*}

We constrain $A_{\rm CB}$ with the SPTpol $B$-mode power spectrum \cite{SPT:2019:BB} since that has the lowest measurement errors at small angular scales. We use the public data that contains the auto and cross spectra of $150$ and $90$ GHz, the bandpower window function, bandpower covariance matrix, and the seven eigenvectors of the beam covariance matrix. We follow \cite{SPT:2019:BB} to analyze the data; we model the theoretical $B$-mode power spectrum that includes contributions from the anisotropic cosmic birefringence, lensing, Galactic dust, and Poisson sources. We then multiply the overall calibration parameters at each frequency and correct the beam error. In total, we have $15$ parameters; $A_{\rm CB}$, the overall lensing amplitude, Galactic dust amplitude, Poisson source amplitude for the three auto and cross spectra, calibration parameters for the two frequencies, and the seven nuisance parameters for the beam error. We add the same Gaussian priors for the dust amplitude, calibration parameters, and beam errors as in \cite{SPT:2019:BB}. We further add a Gaussian prior to the lensing amplitude with the mean of $1$ and the standard deviation of $0.025$ which corresponds to the current best uncertainty on the lensing amplitude by Planck \cite{Carron:2022:NPIPE-lensing}. We use the Markov Chain Monte Carlo (MCMC) package of \href{https://emcee.readthedocs.io/en/stable/}{{\tt emcee}} to fit the bandpowers to the above model. 

Figure \ref{fig:const} shows the results of our constraint on $A_{\rm CB}$. We also show the angular power spectrum from the cosmic birefringence and lensing with parameters corresponding to the best-fit values. We obtain the $2\,\sigma$ constraint as $A_{\rm CB}\times 10^4=1.03^{+0.91}_{-0.97}$. If we increase the standard deviation of the lensing amplitude in the Gaussian prior to $0.1$, the constraint becomes $A_{\rm CB}\times10^4=1.13^{+1.06}_{-1.10}$ at $2\sigma$. Similar to Ref.~\cite{SPT:2019:BB}, even if we decrease or increase the prior on the dust amplitude by $50\%$, the constraint is almost unchanged and is mostly driven by the $150$\,GHz auto power spectrum. 

\section{Summary and Discussion} \label{sec:summary}

We derived the $B$-mode power spectrum induced by anisotropic cosmic birefringence without the thin LSS approximation. We numerically computed the $B$-mode power spectrum and found that the power spectrum is suppressed by an order of magnitude at $\l\alt 10$ and a factor of two at $\l\agt 100$ compared to that with the thin LSS approximation. We also compared the results of P09 and the discrepancy in the behavior of the $B$-mode power spectrum at low-$\l$ would be the ignorance of the spin-operator to the rotation angle which also depends on the line-of-sight direction. 
We constrained the anisotropic birefringence with the SPTpol $B$-mode power spectrum and obtained $A_{\rm CB}\times 10^4=1.03^{+0.91}_{-0.97}\,(2\sigma)$. 

The SPTpol $B$-mode power spectrum slightly prefers a nonzero amplitude of the anisotropic birefringence at approximately $2\sigma$ statistical significance. The results indicate that the SPTpol $B$-mode power spectrum contains unknown systematics that was not characterized in Ref.~\cite{SPT:2019:BB}, and/or suggests new physics. 

For the former possibility, any $E$-to-$B$ leakage could bias the $B$-mode power spectrum. In Ref.~\cite{SPT:2019:BB}, $E$-to-$B$ leakages are mitigated with the measured $EB$ cross-power spectrum. Noise fluctuations in the $EB$ power spectrum could, however, lead to residuals of the leakage in the $B$-mode power spectrum. One of the possible sources of the $E$-to-$B$ leakage that can mimic the effect of anisotropic cosmic birefringence is the miscalibration angle. In Fig.~\ref{fig:const}, we also show the $B$-mode power spectrum induced by the isotropic rotation of $0.63\,$deg which corresponds to the value estimated in Ref.~\cite{SPT:2020:brief}. The $E$-to-$B$ leakage from this rotation effect is simultaneously mitigated by the method described in Ref.~\cite{SPT:2019:BB}, but some residuals might still exist in the $B$-mode power spectrum. In the parameter search, if we replace the $B$-mode power spectrum from the anisotropic cosmic birefringence with that from the isotropic rotation, we obtain $0.84\,$deg for the best-fit value of the isotropic rotation angle. Thus, to explain the slight preference of the nonzero $A_{\rm CB}$ by the residual $B$-modes from isotropic polarization rotation, we need miscalibration-induced $B$-modes comparable to that estimated in Ref.~\cite{SPT:2020:brief}. 

For the latter possibility, we note that $A_{\rm CB}$ has been constrained by directly reconstructing the anisotropic birefringence angle from the correlations between CMB $E$- and $B$-modes at different angular scales \cite{Kamionkowski:2008,Yadav:2012,Namikawa:2016:biref-est}. The best constraint to date on $A_{\rm CB}$ from the reconstruction is an order of magnitude smaller than that obtained in this work \cite{Gluscevic:2012,PB15:rot,BICEP2:2017lpa,Contreras:2017,Namikawa:2020:biref,Gruppuso:2020,Bortolami:2022whx,BK-LoS:2023,Zagatti:2024}. The constraints from the reconstruction, however, have ignored the effect of the time evolution of the anisotropic cosmic birefringence and are currently not possible to directly compare with our result. We leave a further investigation for the consistency of the constraint on anisotropic cosmic birefringence from the different observables for our future work. 


\begin{acknowledgments}
We thank Eiichiro Komatsu and Thomas Crawford for the helpful comments. This work is supported in part by JSPS KAKENHI Grant No. JP20H05859 and No. JP22K03682. Part of this work uses resources of the National Energy Research Scientific Computing Center (NERSC). The Kavli IPMU is supported by World Premier International Research Center Initiative (WPI Initiative), MEXT, Japan. 
\end{acknowledgments}


\appendix

\onecolumngrid

\section{Useful formula} \label{sec:wigner3j}

The spin-weighted spherical harmonics are related to the Wigner $3j$ symbols as \cite{QTAM}
\al{
	\Int{2}{\hatn}{} {}_{s_1}Y_{\l_1m_1}(\hatn){}_{s_2}Y_{\l_2m_2}(\hatn){}_{s_3}Y_{\l_3m_3}(\hatn) 
	= \sqrt{\frac{(2\l_1+1)(2\l_2+1)(2\l_3+1)}{4\pi}}
	\Wjm{\l_1}{\l_2}{\l_3}{-s_1}{-s_2}{-s_3}\Wjm{\l_1}{\l_2}{\l_3}{m_1}{m_2}{m_3} 
	\,. \label{Eq:3times}
}
In particular, we obtain
\al{
    I^{\l L\l_1}_{mM} &\equiv \Int{2}{\hatn}{} [{}_{2}Y^*_{\l_10}(\hatn){}_{2}Y_{\l m}(\hatn)+{}_{-2}Y^*_{\l_10}(\hatn){}_{-2}Y_{\l m}(\hatn)]Y^*_{LM}(\hatn) 
    \notag \\
    &= 2p^+_{\l L\l_1}(-1)^M\sqrt{\frac{(2\l_1+1)(2\l+1)(2L+1)}{4\pi}}
    \Wjm{\l}{L}{\l_1}{-2}{0}{2}\Wjm{\l}{L}{\l_1}{m}{-M}{0}
    \,. 
}
Using
\al{
    \sum_{mM}\Wjm{\l}{L}{\l_1}{m}{-M}{0}\Wjm{\l}{L}{\l_2}{m}{-M}{0} = \frac{1}{2\l_1+1}\delta_{\l_1\l_2}
	\,,
}
we obtain
\al{
    \sum_{mM}I^{\l L\l_1}_{mM}I^{\l L\l_2}_{mM} = 4p^+_{\l L\l_1}\frac{(2\l+1)(2L+1)}{4\pi}\Wjm{\l}{L}{\l_1}{-2}{0}{2}^2\delta_{\l_1\l_2}
    \,. 
}

\twocolumngrid

\bibliographystyle{mybst}
\bibliography{cite}

\providecommand{\href}[2]{#2}\begingroup\raggedright\begin{thebibliography}{10}

\bibitem{Ni:2007:cpr}
W.-T. Ni {\em Prog. Theor. Phys. Suppl.} {\bf 172} (2008) 49--60, \href{http://arxiv.org/abs/0712.4082}{{\tt arXiv:0712.4082}}.

\bibitem{Komatsu:2022:review}
E.~Komatsu {\em Nature Rev. Phys.} {\bf 4} (2022), no.~7 452--469, \href{http://arxiv.org/abs/2202.13919}{{\tt arXiv:2202.13919}}.

\bibitem{Minami:2020:biref}
Y.~Minami and E.~Komatsu {\em \prl} {\bf 125} (2020), no.~22 221301, \href{http://arxiv.org/abs/2011.11254}{{\tt arXiv:2011.11254}}.

\bibitem{Diego-Palazuelos:2022}
P.~Diego-Palazuelos {\em et~al.} {\em \prl} {\bf 128} (2022), no.~9 091302, \href{http://arxiv.org/abs/2201.07682}{{\tt arXiv:2201.07682}}.

\bibitem{Eskilt:2022:biref-freq}
J.~R. Eskilt {\em \aap} {\bf 662} (2022) A10, \href{http://arxiv.org/abs/2201.13347}{{\tt arXiv:2201.13347}}.

\bibitem{Eskilt:2022:biref-const}
J.~R. Eskilt and E.~Komatsu {\em \prd} {\bf 106} (2022), no.~6 063503, \href{http://arxiv.org/abs/2205.13962}{{\tt arXiv:2205.13962}}.

\bibitem{Eskilt:2023:cosmoglobe}
J.~R. Eskilt {\em et~al.} \href{http://arxiv.org/abs/2305.02268}{{\tt arXiv:2305.02268}}.

\bibitem{POLARBEAR:2024vel}
{\bf POLARBEAR}  {\bf Collaboration} , S.~Adachi {\em et~al.} \href{http://arxiv.org/abs/2403.02096}{{\tt arXiv:2403.02096}}.

\bibitem{Carroll:1998:DE}
S.~M. Carroll {\em \prl} {\bf 81} (1998) 3067--3070, \href{http://arxiv.org/abs/astro-ph/9806099}{{\tt astro-ph/9806099}}.

\bibitem{Panda:2010}
S.~Panda, Y.~Sumitomo, and S.~P. Trivedi {\em \prd} {\bf 83} (2011) 083506, \href{http://arxiv.org/abs/1011.5877}{{\tt arXiv:1011.5877}}.

\bibitem{Fujita:2020aqt}
T.~Fujita, Y.~Minami, K.~Murai, and H.~Nakatsuka {\em \prd} {\bf 103} (2021), no.~6 063508, \href{http://arxiv.org/abs/2008.02473}{{\tt arXiv:2008.02473}}.

\bibitem{Fujita:2020ecn}
T.~Fujita, K.~Murai, H.~Nakatsuka, and S.~Tsujikawa {\em \prd} {\bf 103} (2021), no.~4 043509, \href{http://arxiv.org/abs/2011.11894}{{\tt arXiv:2011.11894}}.

\bibitem{Choi:2021aze}
G.~Choi, W.~Lin, L.~Visinelli, and T.~T. Yanagida {\em \prd} {\bf 104} (2021), no.~10 L101302, \href{http://arxiv.org/abs/2106.12602}{{\tt arXiv:2106.12602}}.

\bibitem{Obata:2021}
I.~Obata {\em \jcap} {\bf 09} (2022) 062, \href{http://arxiv.org/abs/2108.02150}{{\tt arXiv:2108.02150}}.

\bibitem{Gasparotto:2022uqo}
S.~Gasparotto and I.~Obata {\em \jcap} {\bf 08} (2022), no.~08 025, \href{http://arxiv.org/abs/2203.09409}{{\tt arXiv:2203.09409}}.

\bibitem{Galaverni:2023}
M.~Galaverni, F.~Finelli, and D.~Paoletti {\em \prd} {\bf 107} (2023), no.~8 083529, \href{http://arxiv.org/abs/2301.07971}{{\tt arXiv:2301.07971}}.

\bibitem{Murai:2022:EDE}
K.~Murai, F.~Naokawa, T.~Namikawa, and E.~Komatsu {\em \prd} {\bf 107} (2023) L041302, \href{http://arxiv.org/abs/2209.07804}{{\tt arXiv:2209.07804}}.

\bibitem{Eskilt:2023:EDE}
J.~R. Eskilt, L.~Herold, E.~Komatsu, K.~Murai, T.~Namikawa, and F.~Naokawa {\em \prl} {\bf 131} (2023), no.~12 121001, \href{http://arxiv.org/abs/2303.15369}{{\tt arXiv:2303.15369}}.

\bibitem{Finelli:2009}
F.~Finelli and M.~Galaverni {\em \prd} {\bf 79} (2009) 063002, \href{http://arxiv.org/abs/0802.4210}{{\tt arXiv:0802.4210}}.

\bibitem{Liu:2016dcg}
G.-C. Liu and K.-W. Ng {\em Phys. Dark Univ.} {\bf 16} (2017) 22--25, \href{http://arxiv.org/abs/1612.02104}{{\tt arXiv:1612.02104}}.

\bibitem{Fedderke:2019:biref}
M.~A. Fedderke, P.~W. Graham, and S.~Rajendran {\em \prd} {\bf 100} (2019) 015040, \href{http://arxiv.org/abs/1903.02666}{{\tt arXiv:1903.02666}}.

\bibitem{Takahashi:2020tqv}
F.~Takahashi and W.~Yin {\em \jcap} {\bf 04} (2021) 007, \href{http://arxiv.org/abs/2012.11576}{{\tt arXiv:2012.11576}}.

\bibitem{Kitajima:2022jzz}
N.~Kitajima, F.~Kozai, F.~Takahashi, and W.~Yin {\em \jcap} {\bf 10} (2022) 043, \href{http://arxiv.org/abs/2205.05083}{{\tt arXiv:2205.05083}}.

\bibitem{Jain:2022jrp}
M.~Jain, R.~Hagimoto, A.~J. Long, and M.~A. Amin {\em \jcap} {\bf 10} (2022) 090, \href{http://arxiv.org/abs/2208.08391}{{\tt arXiv:2208.08391}}.

\bibitem{Gonzalez:2022mcx}
D.~Gonzalez, N.~Kitajima, F.~Takahashi, and W.~Yin {\em \plb} {\bf 843} (11, 2023) 137990, \href{http://arxiv.org/abs/2211.06849}{{\tt arXiv:2211.06849}}.

\bibitem{Myers:2003fd}
R.~C. Myers and M.~Pospelov {\em \prl} {\bf 90} (2003) 211601, \href{http://arxiv.org/abs/hep-ph/0301124}{{\tt hep-ph/0301124}}.

\bibitem{Balaji:2003sw}
K.~R.~S. Balaji, R.~H. Brandenberger, and D.~A. Easson {\em \jcap} {\bf 12} (2003) 008, \href{http://arxiv.org/abs/hep-ph/0310368}{{\tt hep-ph/0310368}}.

\bibitem{Arvanitaki:2009fg}
A.~Arvanitaki, S.~Dimopoulos, S.~Dubovsky, N.~Kaloper, and J.~March-Russell {\em \prd} {\bf 81} (2010) 123530, \href{http://arxiv.org/abs/0905.4720}{{\tt arXiv:0905.4720}}.

\bibitem{Cornelison:2022:BICEP3}
J.~Cornelison {\em et~al.} {\em Proc. SPIE Int. Soc. Opt. Eng.} {\bf 12190} (2022) 121901X, \href{http://arxiv.org/abs/2207.14796}{{\tt arXiv:2207.14796}}.

\bibitem{BICEPArray}
L.~Moncelsi, P.~A.~R. Ade, Z.~Ahmed, M.~Amiri, D.~Barkats, R.~{Basu Thakur}, C.~A. Bischoff, J.~J. Bock, V.~Buza, J.~R. Cheshire, {\em et~al.} {\em Proc. SPIE Int. Soc. Opt. Eng.} {\bf 11453} (2020) 1145314, \href{http://arxiv.org/abs/2012.04047}{{\tt arXiv:2012.04047}}.

\bibitem{SimonsArray}
A.~Suzuki {\em et~al.} {\em J. Low Temp. Phys.} {\bf 184} (2016), no.~3-4 805--810, \href{http://arxiv.org/abs/1512.07299}{{\tt arXiv:1512.07299}}.

\bibitem{SimonsObservatory}
{The Simons Observatory Collaboration} {\em \jcap} {\bf 02} (2019) 056, \href{http://arxiv.org/abs/1808.07445}{{\tt arXiv:1808.07445}}.

\bibitem{CMBS4}
{CMB-S4 Collaboration} {\em \apj} {\bf 926} (2022), no.~1 54, \href{http://arxiv.org/abs/2008.12619}{{\tt arXiv:2008.12619}}.

\bibitem{LiteBIRD}
{LiteBIRD Collaboration} {\em \ptep} (2, 2022) \href{http://arxiv.org/abs/2202.02773}{{\tt arXiv:2202.02773}}.

\bibitem{Sigl:2018:biref-sup}
G.~Sigl and P.~Trivedi {\em {}} (11, 2018) \href{http://arxiv.org/abs/1811.07873}{{\tt arXiv:1811.07873}}.

\bibitem{Sherwin:2021:biref}
B.~D. Sherwin and T.~Namikawa {\em \mnras} {\bf 520} (2021) 3298–3304, \href{http://arxiv.org/abs/2108.09287}{{\tt arXiv:2108.09287}}.

\bibitem{Nakatsuka:2022}
H.~Nakatsuka, T.~Namikawa, and E.~Komatsu {\em \prd} {\bf 105} (2022) 123509, \href{http://arxiv.org/abs/2203.08560}{{\tt arXiv:2203.08560}}.

\bibitem{Liu:2006}
G.-C. Liu, S.~Lee, and K.-W. Ng {\em \prl} {\bf 97} (2006) 161303, \href{http://arxiv.org/abs/astro-ph/0606248}{{\tt astro-ph/0606248}}.

\bibitem{Lee:2013:const}
S.~Lee, G.-C. Liu, and K.-W. Ng {\em \prd} {\bf 89} (2014), no.~6 063010, \href{http://arxiv.org/abs/1307.6298}{{\tt arXiv:1307.6298}}.

\bibitem{Gubitosi:2014:biref-time}
G.~Gubitosi, M.~Martinelli, and L.~Pagano {\em \jcap} {\bf 12} (2014), no.~2014 020, \href{http://arxiv.org/abs/1410.1799}{{\tt arXiv:1410.1799}}.

\bibitem{Lee:2022:pSZ-biref}
N.~Lee, S.~C. Hotinli, and M.~Kamionkowski {\em \prd} {\bf 106} (2022), no.~8 083518, \href{http://arxiv.org/abs/2207.05687}{{\tt arXiv:2207.05687}}.

\bibitem{Namikawa:2023:pSZ}
T.~Namikawa and I.~Obata {\em \prd} {\bf 108} (2023), no.~8 083510, \href{http://arxiv.org/abs/2306.08875}{{\tt arXiv:2306.08875}}.

\bibitem{Carroll:1997:radio}
S.~M. Carroll and G.~B. Field {\em \prl} {\bf 79} (1997) 2394--2397, \href{http://arxiv.org/abs/astro-ph/9704263}{{\tt astro-ph/9704263}}.

\bibitem{Yin:2024:galaxy}
W.~W. Yin, L.~Dai, J.~Huang, L.~Ji, and S.~Ferraro, {\it ``A New Probe of Cosmic Birefringence Using Galaxy Polarization and Shapes''},  2024.

\bibitem{QUaD:2008ado}
E.~Y.~S. Wu {\em et~al.} {\em \prl} {\bf 102} (2009) 161302, \href{http://arxiv.org/abs/0811.0618}{{\tt arXiv:0811.0618}}.

\bibitem{Komatsu:2010:WMAP7}
E.~Komatsu {\em et~al.} {\em \apj} {\bf 192} (2011) 18, \href{http://arxiv.org/abs/1001.4538}{{\tt arXiv:1001.4538}}.

\bibitem{B1rot}
J.~P. Kaufman, N.~J. Miller, M.~Shimon, D.~Barkats, C.~Bischoff, I.~Buder, B.~G. Keating, J.~M. Kovac, {\em et~al.} {\em \prd} {\bf 89} (2014) 062006, \href{http://arxiv.org/abs/1312.7877}{{\tt arXiv:1312.7877}}.

\bibitem{P16:rot}
{{\it Planck} Collaboration} {\em \aap} {\bf 596} (2016) A13, \href{http://arxiv.org/abs/1605.08633}{{\tt arXiv:1605.08633}}.

\bibitem{Monelli:2022:HWP}
M.~Monelli, E.~Komatsu, A.~E. Adler, M.~Billi, P.~Campeti, N.~Dachlythra, A.~J. Duivenvoorden, J.~E. Gudmundsson, and M.~Reinecke \href{http://arxiv.org/abs/2211.05685}{{\tt arXiv:2211.05685}}.

\bibitem{Lue:1999:biref-EB}
A.~Lue, L.~Wang, and M.~Kamionkowski {\em \prl} {\bf 83} (1999) 1506--1509, \href{http://arxiv.org/abs/astro-ph/9812088}{{\tt astro-ph/9812088}}.

\bibitem{Caldwell:2011}
R.~R. Caldwell, V.~Gluscevic, and M.~Kamionkowski {\em \prd} {\bf 84} (2011) 043504, \href{http://arxiv.org/abs/1104.1634}{{\tt arXiv:1104.1634}}.

\bibitem{Lee:2015}
S.~Lee, G.-C. Liu, and K.-W. Ng {\em \plb} {\bf 746} (2015) 406--409, \href{http://arxiv.org/abs/1403.5585}{{\tt arXiv:1403.5585}}.

\bibitem{Leon:2017}
D.~Leon, J.~Kaufman, B.~Keating, and M.~Mewes {\em Mod. Phys. Lett. A} {\bf 32} (2017) 1730002, \href{http://arxiv.org/abs/1611.00418}{{\tt arXiv:1611.00418}}.

\bibitem{Yin:2023:biref}
L.~Yin, J.~Kochappan, T.~Ghosh, and B.-H. Lee {\em \jcap} {\bf 10} (2023) 007, \href{http://arxiv.org/abs/2305.07937}{{\tt arXiv:2305.07937}}.

\bibitem{Ferreira:2023}
R.~Z. Ferreira, S.~Gasparotto, T.~Hiramatsu, I.~Obata, and O.~Pujolas \href{http://arxiv.org/abs/2312.14104}{{\tt arXiv:2312.14104}}.

\bibitem{Li:2008tma}
M.~Li and X.~Zhang {\em \prd} {\bf 78} (2008) 103516, \href{http://arxiv.org/abs/0810.0403}{{\tt arXiv:0810.0403}}.

\bibitem{Li:2013}
M.~Li and B.~Yu {\em \jcap} {\bf 06} (2013) 016, \href{http://arxiv.org/abs/1303.1881}{{\tt arXiv:1303.1881}}.

\bibitem{Zhao:2014}
W.~Zhao and M.~Li {\em \prd} {\bf 89} (2014), no.~10 103518, \href{http://arxiv.org/abs/1403.3997}{{\tt arXiv:1403.3997}}.

\bibitem{Lee:2016}
S.~Lee, G.-C. Liu, and K.-W. Ng {\em The Universe} {\bf 4} (2016), no.~4 29--44, \href{http://arxiv.org/abs/1912.12903}{{\tt arXiv:1912.12903}}.

\bibitem{Pogosian:2019:forecast}
L.~Pogosian, M.~Shimon, M.~Mewes, and B.~Keating {\em \prd} {\bf 100} (2019), no.~2 023507, \href{http://arxiv.org/abs/1904.07855}{{\tt arXiv:1904.07855}}.

\bibitem{Cai:2021:classrot}
H.~Cai and Y.~Guan {\em \prd} {\bf 105} (2022), no.~6 063536, \href{http://arxiv.org/abs/2111.14199}{{\tt arXiv:2111.14199}}.

\bibitem{CoS:2021}
T.~Namikawa, A.~Naruko, R.~Saito, A.~Taruya, and D.~Yamauchi {\em \jcap} {\bf 10} (2021) 029, \href{http://arxiv.org/abs/2103.10639}{{\tt arXiv:2103.10639}}.

\bibitem{Pospelov:2009}
M.~Pospelov, A.~Ritz, and C.~Skordis {\em \prl} {\bf 103} (2009) 051302, \href{http://arxiv.org/abs/0808.0673}{{\tt arXiv:0808.0673}}.

\bibitem{Greco:2024}
A.~Greco, N.~Bartolo, and A.~Gruppuso \href{http://arxiv.org/abs/2401.07079}{{\tt arXiv:2401.07079}}.

\bibitem{Harari:1992}
D.~Harari and P.~Sikivie {\em \prb} {\bf 289} (1992) 67--72.

\bibitem{Carroll:1989:rot}
S.~M. Carroll, G.~B. Field, and R.~Jackiw {\em \prd} {\bf 41} (1990) 1231.

\bibitem{Carroll&Field:1991}
S.~M. Carroll and G.~B. Field {\em \prd} {\bf 43} (Jun, 1991) 3789--3793.

\bibitem{Namikawa:2021:mode}
T.~Namikawa {\em \mnras} {\bf 506} (2021), no.~1 1250--1257, \href{http://arxiv.org/abs/2105.03367}{{\tt arXiv:2105.03367}}.

\bibitem{Naokawa:2023}
F.~Naokawa and T.~Namikawa {\em \prd} {\bf 108} (2023), no.~6 063525, \href{http://arxiv.org/abs/2305.13976}{{\tt arXiv:2305.13976}}.

\bibitem{Lewis:2006:review}
A.~Lewis and A.~Challinor {\em Phys. Rep.} {\bf 429} (2006) 1--65, \href{http://arxiv.org/abs/astro-ph/0601594}{{\tt astro-ph/0601594}}.

\bibitem{Greco:2022:aniso-biref-tomography}
A.~Greco, N.~Bartolo, and A.~Gruppuso {\em \jcap} {\bf 05} (2023) 026, \href{http://arxiv.org/abs/2211.06380}{{\tt arXiv:2211.06380}}.

\bibitem{Zaldarriaga:1996xe}
M.~Zaldarriaga and U.~Seljak {\em \prd} {\bf 55} (1997) 1830--1840, \href{http://arxiv.org/abs/astro-ph/9609170}{{\tt astro-ph/9609170}}.

\bibitem{Pogosian:2011:thick}
L.~Pogosian, A.~P.~S. Yadav, Y.-F. Ng, and T.~Vachaspati {\em \prd} {\bf 84} (2011) 043530, \href{http://arxiv.org/abs/1106.1438}{{\tt arXiv:1106.1438}}. [Erratum: Phys.Rev.D 84, 089903 (2011)].

\bibitem{Hu:1997:TAM}
W.~Hu and M.~J. White {\em \prd} {\bf 56} (1997) 596--615, \href{http://arxiv.org/abs/astro-ph/9702170}{{\tt astro-ph/9702170}}.

\bibitem{CLASS}
D.~{Blas}, J.~{Lesgourgues}, and T.~{Tram} {\em \jcap} {\bf 2011} (2011), no.~7 034, \href{http://arxiv.org/abs/1104.2933}{{\tt arXiv:1104.2933}}.

\bibitem{Namikawa:2020:biref}
T.~Namikawa {\em et~al.} {\em \prd} {\bf 101} (2020), no.~8 083527, \href{http://arxiv.org/abs/2001.10465}{{\tt arXiv:2001.10465}}.

\bibitem{SPT:2020:brief}
F.~Bianchini {\em et~al.} {\em \prd} {\bf 102} (2020), no.~8 083504, \href{http://arxiv.org/abs/2006.08061}{{\tt arXiv:2006.08061}}.

\bibitem{Zagatti:2024}
G.~Zagatti, M.~Bortolami, A.~Gruppuso, P.~Natoli, L.~Pagano, and G.~Fabbian \href{http://arxiv.org/abs/2401.11973}{{\tt arXiv:2401.11973}}.

\bibitem{SPT:2019:BB}
J.~T. Sayre {\em et~al.} {\em \prd} {\bf 101} (2020), no.~12 122003, \href{http://arxiv.org/abs/1910.05748}{{\tt arXiv:1910.05748}}.

\bibitem{Carron:2022:NPIPE-lensing}
J.~Carron, M.~Mirmelstein, and A.~Lewis {\em \jcap} {\bf 09} (2022) 039, \href{http://arxiv.org/abs/2206.07773}{{\tt arXiv:2206.07773}}.

\bibitem{Kamionkowski:2008}
M.~Kamionkowski {\em \prl} {\bf 102} (2009) 111302, \href{http://arxiv.org/abs/0810.1286}{{\tt arXiv:0810.1286}}.

\bibitem{Yadav:2012}
A.~P.~S. {Yadav}, M.~{Shimon}, and B.~G. {Keating} {\em \prd} {\bf 86} (2012), no.~8 083002, \href{http://arxiv.org/abs/1207.6640}{{\tt arXiv:1207.6640}}.

\bibitem{Namikawa:2016:biref-est}
T.~Namikawa {\em \prd} {\bf 95} (2017), no.~4 043523, \href{http://arxiv.org/abs/1612.07855}{{\tt arXiv:1612.07855}}.

\bibitem{Gluscevic:2012}
V.~{Gluscevic}, D.~{Hanson}, M.~{Kamionkowski}, and C.~M. {Hirata} {\em \prd} {\bf 86} (2012) 103529, \href{http://arxiv.org/abs/1206.5546}{{\tt arXiv:1206.5546}}.

\bibitem{PB15:rot}
{\textsc{POLARBEAR} Collaboration} {\em \prd} {\bf 92} (2015) 123509, \href{http://arxiv.org/abs/1509.02461}{{\tt arXiv:1509.02461}}.

\bibitem{BICEP2:2017lpa}
{BICEP2 and Keck Array Collaborations} {\em \prd} {\bf 96} (2017), no.~10 102003, \href{http://arxiv.org/abs/1705.02523}{{\tt arXiv:1705.02523}}.

\bibitem{Contreras:2017}
D.~Contreras, P.~Boubel, and D.~Scott {\em \jcap} {\bf 1712} (2017), no.~12 046, \href{http://arxiv.org/abs/1705.06387}{{\tt arXiv:1705.06387}}.

\bibitem{Gruppuso:2020}
A.~Gruppuso, D.~Molinari, P.~Natoli, and L.~Pagano {\em \jcap} {\bf 11} (2020) 066, \href{http://arxiv.org/abs/2008.10334}{{\tt arXiv:2008.10334}}.

\bibitem{Bortolami:2022whx}
M.~Bortolami, M.~Billi, A.~Gruppuso, P.~Natoli, and L.~Pagano {\em \jcap} {\bf 09} (2022) 075, \href{http://arxiv.org/abs/2206.01635}{{\tt arXiv:2206.01635}}.

\bibitem{BK-LoS:2023}
{BICEP2 and Keck Array Collaborations} {\em \apj} {\bf 949} (2023), no.~2 43, \href{http://arxiv.org/abs/2210.08038}{{\tt arXiv:2210.08038}}.

\bibitem{QTAM}
D.~Varshalovich, A.~Moskalev, and V.~Kersonskii, {\em Quantum Theory of Angular Momentum}.
\newblock World Scientific, 1989.

\end{thebibliography}\endgroup

\end{document}